\documentclass{PoS}

\usepackage{amsmath}
\usepackage{epsfig}

\title{The dependence of the number of pomerons on the impact parameter and the long-range
rapidity correlations in pp collisions}

\ShortTitle{The dependence of the number of pomerons on the impact parameter and LRC in pp}

\author{\speaker{Vladimir Vechernin}%
         \thanks{The work was supported by the RFFI grant 12-02-00356-a.}\\
        Saint-Petersburg State University\\
        E-mail: \email{vechernin@gmail.com}}

\author{Igor Lakomov\\
        Saint-Petersburg State University\\
        E-mail: \email{lakomov@gmail.com}}

\abstract{
The simple model which enables to take into
account the effects of a colour string fusion in $pp$ interactions is suggested.
The parameters of the model are connected
with  the parameters of the pomeron trajectory and its couplings to hadrons.
On the base of the model the MC algorithm which enables to calculate
the long-range correlation functions between multiplicities and
between the average transverse momentum and the multiplicity in \textit{pp}
collisions is developed.
}
\FullConference{XXI International Baldin Seminar on High Energy Physics Problems \\
                 September 10-15, 2012\\
                 JINR, Dubna, Russia}

\def\bc{\begin{center}}
\def\ec{\end{center}}
\def\beq{\begin{equation}}
\def\eeq{\end{equation}}
\def\beq*{\begin{equation*}}
\def\eeq*{\end{equation*}}

\def\hs#1{\hspace*{#1cm}}
\def\av#1{\langle #1 \rangle}
\def\avr#1#2{\langle {#1} \rangle^{}_{#2}}
\def\Nb{\overline{N}}
\def\DF{{\Delta y^{}_F}}
\def\DB{{\Delta y^{}_B}}
\def\Dy{{\Delta y}}
\def\Deta{{\Delta \eta}}

\def\nF{{n_F^{}}}
\def\pF{{p_{tF}^{}}}
\def\nB{{n_B^{}}}
\def\pB{{p_{tB}^{}}}

\begin{document}

\section{Introduction}
The soft part of the multi-particle production
in pp collisions at high energy,
which can not be directly described in terms of the perturbative QCD,
is usually described in the framework of string models \cite{Capella0}--\cite{Kaidalov1},
which originate from the Gribov-Regge approach.
In these models at first stage the colour strings are stretched between
the projectile and target partons. The formation of the pair of strings
corresponds to the cut pomeron in the Gribov-Regge approach.
The ha\-d\-ro\-ni\-za\-tion of these strings produces the observed hadrons.

 In the case of nuclear collisions, the number of strings grows
with the growing energy and the number of nucleons of colliding nuclei, and
one has to take into account the interaction between
strings in the form of their fusion and/or percolation
\cite{BP1}--\cite{PRLperc}.
The fusion process results in the reduction
of total multiplicity of charged particles and
growth of transverse momentum, that
was confirmed later \cite{comparRHIC1,comparRHIC2}
in comparison with RHIC data.

The possible experimental observation of the string fusion or percolation
 is extremely interesting. Therefore, the investigation of long-range
correlations (LRC) between multiplicities and transverse momenta in two separated rapidity intervals
were proposed as the main tool to study this
phenomenon \cite{PRL94}--\cite{ALP}.

Up to the present the string fusion model was usually used for describing
the multiplicity and mean transverse momentum of charged
particles and their correlation only in the case of
nucleus-nucleus collisions. However, the experimental data indicates
an increase with energy of the mean transverse momentum and its correlation with multiplicity
also in $pp$ collisions \cite{pp1}--\cite{pp5}.
In present work  we formulate the simple model which enables to take into account
the effect of colour string fusion on the LRC between multiplicities and transverse momenta
in $pp$ interactions.

\section{Formulation of the model}
To take into account  the effect of string fusion  in $pp$ collisions
one needs to know the distribution of strings in the transverse plane
at given value of the impact parameter $b$.
We'll do this in an analogy with the case of nucleus-nucleus collisions.

\subsection{Distribution of strings in the transverse plane: $AA$ interactions}
In the case of $AA$ interactions one usually assumes
that at high energy a number of primary formed quark-gluon strings
is proportional to the number of binary inelastic nucleon-nucleon
interactions in a given event of nucleus-nucleus scattering \cite{Capella1,BPep00}.
In frame of the classical Glauber model mean number of
inelastic $NN$ interactions in $AB$ scattering at a fixed impact
parameter $b$ is given by the expression (e.g. see \cite{Bialas76,PRC11}):
\begin{equation}
\label{a3}
\langle N_{str}(b)\rangle \sim \langle N^{in}_{coll}(b)\rangle = A B
\frac{\sigma_{\!N\!N}^{in}}{\sigma_{\!AB}^{}(b)}
\int{T_{A}(\vec{s}-\vec{b}/2)T_{B}(\vec{s}+\vec{b}/2)d^{2}\vec{s}}  \ .
\end{equation}
Here $\sigma_{\!N\!N}^{in}$ -- the cross-section of inelastic
nucleon-nucleon interaction, $\sigma_{\!AB}^{}(b)$ -- the probability
of interaction of two nuclei
at a given impact parameter
$b$ with at least one inelastic $NN$ interaction.
An integral of $\sigma_{\!AB}^{}(b)$:
\begin{equation}
\label{a3a} \sigma_{\!AB}^{} = \int \sigma_{\!AB}^{}(b)\ d^{2}\vec{b}
\end{equation}
gives a so-called <<production cross section>> of $AB$ interaction.
The $\langle...\rangle$ in equation (\ref{a3})
means an averaging over all events with a given $b$.
$T_{A}$ and $T_{B}$ -- the profile functions of the colliding nuclei:
\begin{equation}
\label{a2} T_{A}(\vec{s}) = \int\limits_{-\infty}^{+\infty}{\rho_{A}(\vec{s}, z) dz} \ ,
\end{equation}
where $\rho_{A}(\vec{r})$ -- the nucleon density of nucleus $A$, normalized to unity,
$\vec{r}=(\vec{s}, z)$. The $\vec{s}$ is a two-dimensional vector
in the impact parameter plane.

By (\ref{a3})
in the nucleus-nucleus collision at the impact parameter $b$
for the string density in transverse plane at a point $\vec{s}$
we have:
\begin{equation}
\label{a4}
d\langle N_{str}(b)\rangle /d^2\vec{s}\equiv w_{str}(\vec{s},\vec{b})\sim w^{in}_{coll}(\vec{s},\vec{b})
= A B
\frac{\sigma_{\!N\!N}^{in}}{\sigma_{\!AB}^{}(b)}T_{A}(\vec{s}-\vec{b}/2)T_{B}(\vec{s}+\vec{b}/2)
\end{equation}
Note that in the transverse plane we place the origin in the middle between the centers of colliding nuclei.

\subsection{Distribution of strings in the transverse plane: $pp$ interactions}
By (\ref{a4}) it is natural to suppose that  in the case of  proton-proton collision at the impact parameter $b$
the string density in transverse plane at a point $\vec{s}$ is proportional to
\begin{equation}
\label{a9a}
w_{str}(\vec{s}, \vec{b}) \sim
T(\vec{s}-\vec{b}/2)T(\vec{s}+\vec{b}/2)/\sigma_{\!pp}^{}(b)   \ ,
\end{equation}
where now the $T(\vec{s})$  is the partonic profile function of nucleon.
The $\sigma_{\!pp}^{}(b)$ is the probability of non-diffractive $pp$ interaction
(with at least one cut pomeron) at a given impact parameter $b$. An
integral of $\sigma_{\!pp}^{}(b)$:
\begin{equation}
\label{a3b} \sigma_{\!pp}^{} = \int \sigma_{\!pp}^{}(b)\ d^{2}\vec{b}
\end{equation}
gives a non-diffractive cross-section of $pp$ interaction.

By analogy with light nuclei we will use
for the partonic profile function of nucleon
the simplest gaussian distribution:
\begin{equation}
\label{a6}
T(s) =
\frac
{e^{-s^2/\alpha^2}}
{\pi\alpha^2}\ .
\end{equation}
Substituting (\ref{a6}) in (\ref{a9a}) one gets
\begin{equation}
\label{factor}
w_{str}(\vec{s}, \vec{b}) \sim
e^{-(\vec{s}+\vec{b}/2)^2/\alpha^2}
e^{-(\vec{s}-\vec{b}/2)^2/\alpha^2}/\sigma_{\!pp}^{}(b)=
e^{-2 s^2 / \alpha^2}
e^{-b^2 / 2\alpha^2}/\sigma_{\!pp}^{}(b)     \ .
\end{equation}
We see that in this approximation the dependencies on $b$ and $s$ are factorized and after integration on $\vec{s}$
one gets for the mean number of strings in the $pp$ collision at the impact parameter $b$:
\begin{equation}
\label{a90}
\langle N_{str}(b) \rangle
 \sim e^{-b^2 / 2\alpha^2}/\sigma_{\!pp}^{}(b)   \ .
\end{equation}

Since in this approach the formation of each pair of strings
corresponds to one cut pomeron, $N_{str}=2N$,  where  $N$  is the number of cut pomerons in a given event,
hence (\ref{a90}) leads to
\begin{equation}
\label{cutpom}
\langle N(b) \rangle
\sim e^{-b^2 / 2\alpha^2}/\sigma_{\!pp}^{}(b)   \ ,
\end{equation}
which gives
the dependence of the average number of cut pomerons on the
impact parameter in non-diffractive \textit{pp} collisions.

\subsection{Event-by-event fluctuations of the number of cut pomerons at given impact parameter}
For the calculation of LRC one should know not only the mean
number of pomerons in $pp$ collisions at a given impact parameter $b$,
but also the event by event distribution of the number of pomerons around this mean value.
We assume that this distribution $\widetilde{P}(N,b)$
at a given value of the impact parameter $b$ at $N\geq 1$:
\begin{equation}
\label{a10a} \widetilde{P}(N,b) = {P(N,b)}/ [1 - P(0,b)]
\end{equation}
is the simple modification of the poissonian distribution:
\begin{equation}
\label{a10} P(N,b) = e^{-\overline{N}(b)} {\overline{N}(b)^{N}}/ {N!}
\end{equation}
with some parameter $\overline{N}(b)$. The difference of our
distribution $\widetilde{P}(N,b)$ (\ref{a10a}) from the poissonian
one (\ref{a10}) is only in excluding of the point $N=0$ from it: $\widetilde{P}(0,b)=0$,
which corresponds to the absence of the non-diffractive scattering at $N=0$.
Clear that at $N\geq 1$ this reduces only to the introduction of
the additional common normalization factor in (\ref{a10a}),
which enables the proper normalization: $\sum_{N=1}\widetilde{P}(N,b) =1$.

The calculation of the mean number of pomerons at a given $b$ with
 the distribution (\ref{a10a}) gives:
\begin{equation}
\label{a10b} \langle N(b) \rangle = {\overline{N}(b)}/ [1 - P(0,b)] \ .
\end{equation}

Since the probability $\sigma_{\!pp}^{}(b)$ of the non-diffractive $pp$
interaction at the given fixed impact parameter $b$ is equal to
the probability to have a nonzero number of cut pomerons, then
\begin{equation}
\label{pp} \sigma_{\!pp}^{}(b)=1 - P(0,b)=1 - \exp(-\overline{N}(b)) \ .
\end{equation}
Comparing now the formulae (\ref{cutpom}) and (\ref{a10b}) with
taking into account (\ref{pp}) we see that in our model $\overline{N}(b) \sim e^{-b^2 / 2\alpha^2}$ or
introducing a parameter $N_0$:
\begin{equation}
\label{a12} \overline{N}(b) =N_0 e^{-b^2 / 2\alpha^2}
\end{equation}
Then the mean number of cut pomerons at an impact parameter $b$ is given by
\begin{equation}
\label{a12a} \langle N(b) \rangle = {\overline{N}(b)}/ [1 - \exp(-\overline{N}(b))] \ ,
\end{equation}
where the $\overline{N}(b)$ is given by (\ref{a12}).

\subsection{{Integration over impact parameter - min.bias \textit{pp} collisions}}

It is convenient to introduce in the impact parameter plane a density of the probability of
 non-diffractive $pp$ interaction (see (\ref{a3b})) normalized to unity:
\begin{equation}
\label{a11} f(b) = \sigma_{\!pp}^{}(b)/\sigma_{\!pp}^{}  \ , \hs 2 \int f(b)\, d^{2}\vec{b} =1 \ .
\end{equation}
Using the formula (\ref{a11}) one can find a mean number of pomerons in
non-diffractive $pp$ interaction  averaged  over the impact parameter:
\begin{equation}
\label{a13} \langle N\rangle = \int \langle N(b) \rangle\, f(b)\, d^{2}\vec{b} = \int \overline{N}(b)\,
d^{2}\vec{b}/\sigma_{\!pp}^{} = 2\pi \alpha^2 N_0 /\sigma_{\!pp}^{}
\end{equation}
and the corresponding variance $D_N \equiv \langle N^2 \rangle  -  \langle N^{} \rangle^2_{}$, where
\begin{equation}
\label{a13a2}
 \langle N^2 \rangle
= \int \langle N^2(b) \rangle\, f(b)\, d^{2}\vec{b}
= \pi \alpha^2 N_0 (N_0+2)/\sigma_{\!pp}^{}
= \langle N\rangle\, (N_0+2)/2
\end{equation}
Here we have used  (\ref{a12}). In our model using (\ref{a12}) we can calculate also the cross-section of non-diffractive $pp$ interaction
\begin{equation}
\label{a13s}
\sigma_{\!pp}^{} =
\int \sigma_{\!pp}^{}(b)\,d^2 \vec{b}=
\int [1 - \exp(-\overline{N}(b))]\,d^2 \vec{b}=  2\pi\alpha^{2}  \Phi_1(N_0)  \ ,
\end{equation}
where
\begin{equation}
\label{spec}
\Phi_1(z)=\int_0^z(1-e^{-t})\frac{dt}{t}
 \ , \hs{0.5}
\Phi_m(z)=\sum_{k=1}^{\infty}\frac{(-1)^{k+1}z^k}{k!k^m}
  \ .
\end{equation}

\subsection{Probability of $N$ cut pomerons in non-diffractive \textit{pp} collision}
In the framework of our simple model we can find  the probability $w^{}_N$ to have $N$ cut pomerons in a non-diffractive \textit{pp} collision
by averaging the $\widetilde{P}(N,b)$  (\ref{a10a}) over $b$ at fixed $N$:
\begin{equation}
\label{wN}
w^{}_N
= \int \widetilde{P}(N,b)\, f(b)\, d^{2}\vec{b}
= \frac{1}{\sigma_{\!pp}^{}} \int P(N,b)\, d^{2}\vec{b} \ ,
\end{equation}
where we have taken into account (\ref{pp}). Using now (\ref{a10}), we have
\begin{equation}
\label{wNint}
w_N = \frac{1}{\sigma_{\!pp}^{} {N!}} \int   e^{-\overline{N}(b)}
(\overline{N}(b))_{}^{N} \, d^{2}\vec{b} =
\frac{2\pi}{\sigma_{\!pp}^{} {N!}} \int_0^\infty   e^{-\overline{N}(b)}
(\overline{N}(b))_{}^{N} \, b\,db  \ .
\end{equation}
One can introduce in (\ref{wNint}) the new integration variable $\Nb$ in accordance with (\ref{a12}):
\begin{equation}
\label{Nbar}
\Nb=\Nb(b)=N_0 e^{-b^2 / 2\alpha^2} \ , \hs1
d\Nb=-(\Nb/\alpha^2)  \   b\,db  \ .
\end{equation}
Then  (\ref{wNint}) takes the following form
\begin{equation}
\label{wNint1}
w_N =
\frac{2\pi\alpha^2}{\sigma_{\!pp}^{} {N!}}
\int_0^{N_0}   e^{-\Nb} \Nb_{}^{N-1} \, d\Nb  \ .
\end{equation}
Such integral  is a difference of the gamma  and  incomplete gamma functions:
\begin{equation}
\label{intz}
\int_0^{N_0}   e^{-z} z_{}^{N-1} \, dz=\int_0^{\infty}   e^{-z} z_{}^{N-1} \, dz
-\int_{N_0}^\infty   e^{-z} z_{}^{N-1} \, dz
=\Gamma(N)-\Gamma(N,N_0)  \ ,
\end{equation}
At integer $N$
\begin{equation}
\label{Gam}
\Gamma(N)=(N-1)!  \ , \hs1
\Gamma(N,N_0) =(N-1)! \, e^{-N_0} \sum_{l=0}^{N-1} N_0^l/l! \ .
\end{equation}
Gathering we find
\begin{equation}
\label{w_Nr}
w_N =
\frac{2\pi\alpha^2}{\sigma_{\!pp}^{} N}
\left[1- e^{-N_0} \sum_{l=0}^{N-1} N_0^l/l! \right] =\frac{\sigma^{}_N}{\sigma_{\!pp}^{}}    \ ,
\end{equation}
where we have introduced the $\sigma^{}_N$ by
\begin{equation}
\label{sigN}
\sigma_{N}^{} \equiv
 \frac{2\pi\alpha^{2}}{N}\left[1- e^{-N_0} \sum_{l=0}^{N-1} N_0^l/l! \right] \ ,
\end{equation}
The direct summing gives
\begin{equation}
\label{sigpp}
\sum_{N=1}^{\infty} \sigma_{N}^{}
 = 2\pi\alpha^{2}\Phi_1(N_0)=\sigma_{\!pp}^{} \ ,
\end{equation}
where we have used (\ref{a13s}) and (\ref{spec}). Recall that $\sigma_{\!pp}^{}$ is the non-diffractive $pp$ cross section  (\ref{a3b}).

\subsection{Comparison with quasi-eikonal and Regge approaches}
Now we see that our formula for  the $\sigma^{}_N$ (\ref{sigN}) coincides with
the well known result for the cross-section $\sigma^{}_N$ of $N$ cut-pomeron exchange,
obtained in the quasi-eikonal and Regge approaches  \cite{Ter-Mart73,AGK,Kaid84} :
\begin{equation}
\label{R}
\sigma^{}_N =
\frac{4\pi\lambda}{C\,N} \left[1- e^{-z} \sum_{k=0}^{N-1} z^k/k! \right] \ ,
\end{equation}
where
\begin{equation}
\label{Rpar}
z=\frac{2\gamma C}{\lambda} \exp(\Delta\xi) \ , \hs{0.5}
\lambda=R^2+\alpha'\xi   \ , \hs{0.5}
\xi=\ln (s/1 GeV^2) \ .
\end{equation}
Here  $\Delta$ and $\alpha'$ are the residue and the slope of the
pomeron trajectory.
The parameters $\gamma$ and $R$ characterize the coupling of the pomeron trajectory
with the initial hadrons. The quasi-eikonal parameter $C$ is related to
the small-mass diffraction dissociation of incoming hadrons.

This enables to connect  the parameters $N_0$ and $\alpha$
of our model, which describe the dependence of
the mean number of pomerons on the impact parameter $b$
(see formulae (\ref{a12}) and (\ref{a12a}))
with  the parameters of the pomeron trajectory
and its couplings to hadrons. Comparing (\ref{sigN}) and (\ref{R}) we have
\begin{equation}
\label{paramfix}
N_0=z=\frac{2\gamma C}{\lambda} \exp(\Delta\xi)   \ , \hs{0.5}
\alpha=\sqrt{\frac{2\lambda}{C}}  \ , \hs{0.5}
\lambda=R^2+\alpha'\xi
\end{equation}
In our calculations the numerical values of the parameters for the case of $pp$ collisions
were taken from the paper \cite{KaidShab02}:
\begin{equation}
\label{paramnum}
\Delta= 0.139,\ \ \ \alpha'=0.21\ GeV^{-2},\ \ \ \gamma=1.77\ GeV^{-2},\ \ \
R^2=3.18\ GeV^{-2},\ \ \ C=1.5\ ,
\end{equation}
which gives, for example, for the parameters $N_0$ and $\alpha$ entering in our formula (\ref{a12}):
$$
\alpha=0.51\textrm{fm},\ N_0=3.38 \ \ \  \textrm{at}\ \sqrt{s}=60GeV \ ,
$$
\begin{equation}
\label{paramour}
\alpha=0.60\textrm{fm},\ N_0=9.02 \ \ \  \textrm{at}\ \sqrt{s}=7000GeV \ .
\end{equation}

\section{String fusion effects}
As follows from the consideration in the previous section,
in $pp$ interactions the number of contributing cut pomerons
(see formulae (\ref{a12}) and (\ref{a12a}))
increases with the collision centrality and energy (\ref{paramour}).
In our approach the number of strings is twice the number of cut pomerons,
so it also will increases with the collision centrality and energy.
Since the strings have a certain limited size in the
transverse plane (a plane of the impact parameter) an overlap of
the strings will start with increase of their density. As
a result the color fields of different strings will interact what
will influence on their fragmentation process. For taking into account the effects
from the interaction of strings at high density a string fusion model (SFM) was proposed
\cite{BP1}--\cite{PRLperc}.

\subsection{Different versions of string fusion model}
There are two versions of the SFM a model with local fusion (overlaps model) \cite{loc} and
a model with formation of global color clusters (clusters model)
\cite{glob}.  In both versions the fusion processes result in the reduction
of total multiplicity of charged particles and
growth of transverse momentum.

In the first variant according to
\cite{loc} we assume that the mean multiplicity per unit of rapidity and
the mean transverse momentum of charged particles emitted from
the region, where $k$ strings are overlapping, are described by the following expressions:
\begin{equation}
\label{loc}
\langle n \rangle_k=\mu_0\sqrt{k}\ S_k/\sigma_0 \ ,
\hskip 1cm
\langle p^2_t \rangle_k=p_0^2 \sqrt{k} \ ,
\hskip 1cm
k=1,2,3,... \ .
\end{equation}
Here $S_k$ is the transverse area of the region, where $k$ color strings are
overlapped. $\sigma_0=\pi r^2_{str}$ is the transverse area of a string. $\mu_0$ and
$p_0$ are the mean multiplicity per unit of rapidity and the mean
transverse momentum of charged particles, produced from a decay of one
string.

\begin{figure}
\centering
\includegraphics[width=0.8\linewidth]{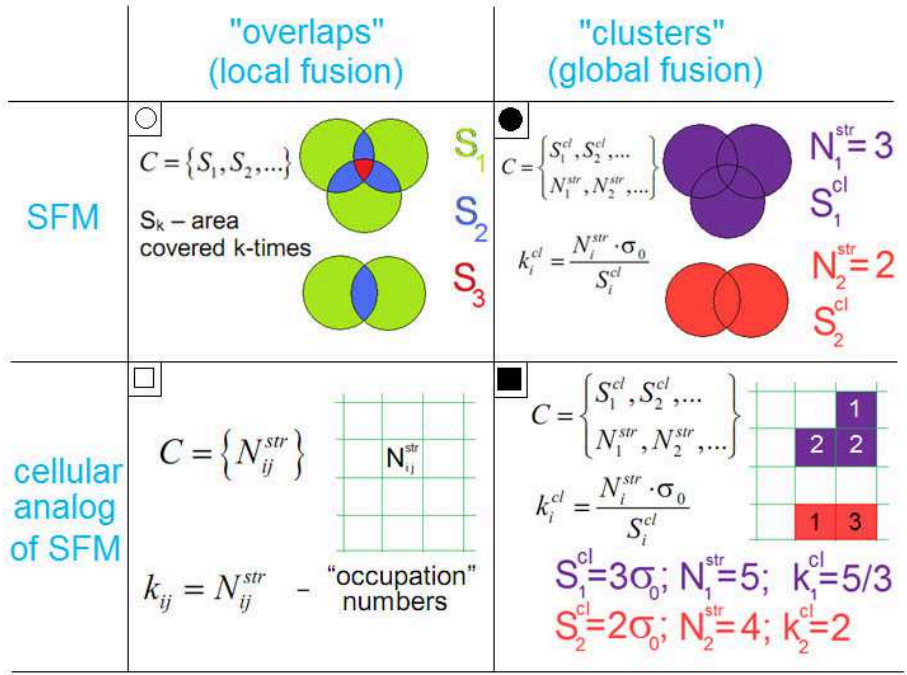}
\caption{The different versions of the SFM.
The first column corresponds to the version with a local string fusion (overlaps model).
The second column corresponds to the formation of fused string clusters (clusters model).
The first row corresponds to the original versions of the SFM \cite{loc,glob} and
the second row corresponds to their cellular analogs \cite{V1,V2,YF1,YF2} (see text for details).
}
\label{SFMver}
\end{figure}

In the second variant one
assumes that the fused strings form a cluster.
According to \cite{glob} in this case we suppose that the mean multiplicity per unit of
rapidity and the mean transverse momentum of charged particles emitted
by a cluster with a transverse area $S_{cl}$, formed by $k$ strings,
can be found as follows
\begin{equation}
\label{glob}
\langle n \rangle_{cl}=\mu_0\sqrt{k_{cl}}S_{cl}/\sigma_0 \ ,
\hskip 1cm
\langle p_t^2 \rangle_{cl}=p_0^2\sqrt{k_{cl}} \ ,
\hskip 1cm
k_{cl}=k\sigma_0/S_{cl} \ .
\end{equation}
Note that in two limiting cases (an absence of overlaps of
strings and a total overlapping of string areas)
both variants lead to the same rules.

Later the simplified discrete analogs of both mentioned versions of SFM
based on the implementation of the lattice in the transverse plane
\cite{V1,V2,EPJC04} were proposed.  It was demonstrated \cite{YF1,YF2} that the MC
algorithms of the calculations of LRC coefficients based on the
cellular versions of SFM work much faster and give
practically the same results as the ones based on the original versions of SFM.
These cellular versions of SFM enable also to calculate analytically the mean
values of the observables and LRC coefficients in some limiting
cases \cite{V2,YF2}, what was used to control the reliability of the created MC codes.

In Fig.\ref{SFMver} we illustrate visually the prescriptions of all these versions of the SFM.
The first column corresponds to the version with a local string fusion (overlaps model).
The second column corresponds to the formation of fused string clusters (clusters model).
The first row corresponds to the original versions of the SFM and
the second row corresponds to their cellular analogs.

\subsection{Monte-Carlo algorithm for the calculation of LRC functions}
The calculation of LRC functions  (regressions) is based on the formula
\begin{equation}
\label{nncor}
\avr\nB\nF=\frac{\sum_C w(C)\   \avr\nB C \ P_C(\nF)}
{\sum_C w(C)\ P_C(\nF)}
\end{equation}
obtained in \cite{BPV00} for the correlation between multiplicities $\nF$ and $\nB$ in separated rapidity
windows and on the similar formula
\begin{equation}
\label{ptncor}
\avr\pB\nF=\frac{\sum_C w(C)\ \avr\pB C \ P_C(\nF)}
{\sum_C w(C)\ P_C(\nF)}
\end{equation}
used in \cite{V2,EPJC04,YF1} for the calculations of the LRC function
between the multiplicity $\nF$ in the forward rapidity window and the
corresponding mean transverse momentum $\pB$ of $\nB$ charged particles in the backward window.

The calculations of the sums on string configurations $C$
in numerator and denominator of the formulae (\ref{nncor}) and (\ref{ptncor})
were performed by MC simulations of the configurations $C$ with proper weights $w(C)$:
\begin{equation}
\label{sim}
\sum_C w(C)\ ... =\frac{1}{n_{sim}} \sum_{sim}\ ...       \ .
\end{equation}
With this purpose at the generation of the string configurations $C$ the results of the Section 2 were used.

For the generated string configuration $C$ the mean values of
the multiplicity $\avr\nB C$ and transverse momentum  $\avr\pB C$
in the backward rapidity window,
entering the formulae (\ref{nncor}) and (\ref{ptncor}),
were calculated
using the prescriptions (\ref{loc}) and (\ref{glob})
for the cases of  local string fusion and cluster formation,
both in frameworks of the original version of SFM and its cellular analog.

We also have supposed that the multiplicity distribution
of charged particles produced from the decay of any string
is  poissonian, what leads to the  poissonian distribution for
 the probability of production of $\nF$ charged particles  $P_C(\nF)$
from the given string configuration $C$:
\begin{equation}
\label{PC}
P_C(\nF)=P_{\avr\nF C}(\nF)
=e^{-{\avr\nF C}}\,\frac{(\avr\nF C)_{}^\nF}{\nF!}  \ ,
\end{equation}
where one can calculate  $\avr\nF C$ similar to $\avr\nB C$
using the prescriptions (\ref{loc}) and (\ref{glob}).

Similarly to  the correlation functions (\ref{nncor}) and (\ref{ptncor})
one can also calculate (by MC simulations of the string configurations $C$)
the overall mean values (averaged over all events) in the forward
\begin{equation}
\label{nF}
\av\nF =\sum_C w(C) \avr\nF C \ ,
\hskip 1cm
\av\pF =\sum_C w(C) \avr\pF C
\end{equation}
and (by the same way) in the backward windows.

Note that the resulting distribution on the number
of charged particles produced in the forward or backward
window
\begin{equation}
\label{PF}
P(\nF) =\sum_C w(C) P_C(\nF)
\end{equation}
will be non-poissonian despite the poissonian form (\ref{PC}) of the distribution $P_C(\nF)$ and
almost poissonian (see (\ref{a10a}) and (\ref{a10})) fluctuations in the number of cut pomerons
at fixed value of the impact parameter $b$ in our model.
The reason is the non-poissonian fluctuations in the number of cut pomerons
originating from the event-by-event fluctuation of the impact parameter $b$ (see also \cite{YF1,Dub10,CERN12}).

\section{Results}
\begin{figure}
\centering
\psfig{file=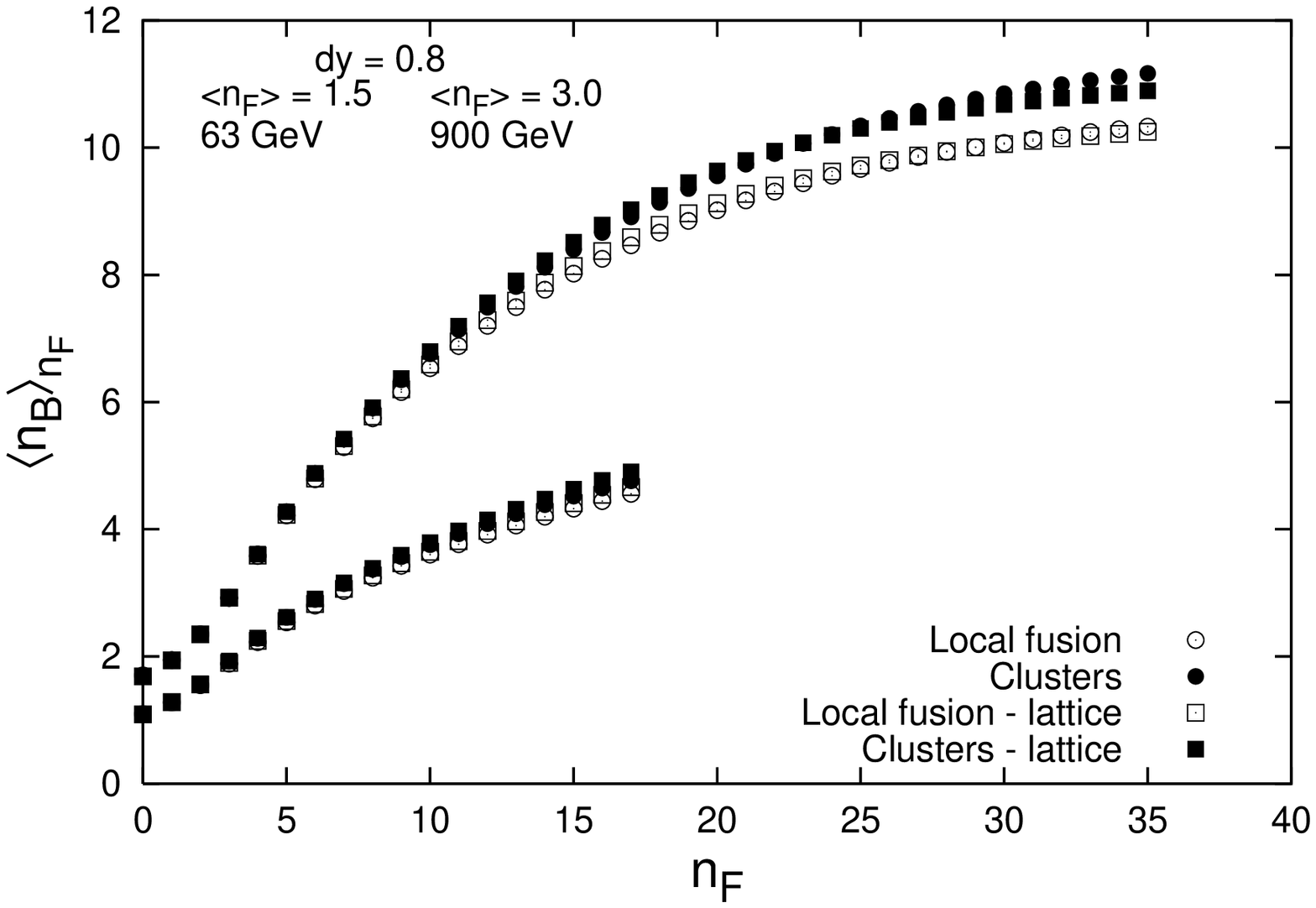,width=8cm,angle=-90}
\caption[dummy]{
The long-range correlation (LRC) function in $pp$ collisions between multiplicities $\nF$ and $\nB$ in separated rapidity
windows calculated by (\ref{nncor}) at energies 63 and 900 GeV for the forward and backward rapidity windows
of the same width $\DF=\DB=dy=0.8$, what corresponds to the mean multiplicities $\av\nF=\av\nB=\av n$=1.5 and 3.0 per
window at these energies.
Points {\Large $\circ$} and {\Large $\bullet$} - the results of calculations in the framework of the
original versions of the string fusion model (SFM) with the local fusion (overlaps) or the formation of string clusters (clusters)
correspondingly (see first row in Fig.\ref{SFMver}) \cite{loc,glob}.
Points \raisebox{1.5mm}{\framebox[2.0mm][c]{}} and $\rule[0.4mm]{2.1mm}{2.1mm}$ -~the results
of calculations in the framework of the lattice analog of SFM
with the local string fusion or the cluster formation
correspondingly (see second row in Fig.\ref{SFMver}) \cite{V1,V2,YF1,YF2}.
}
\label{fignn}
\end{figure}
\begin{figure}
\centering
\psfig{file=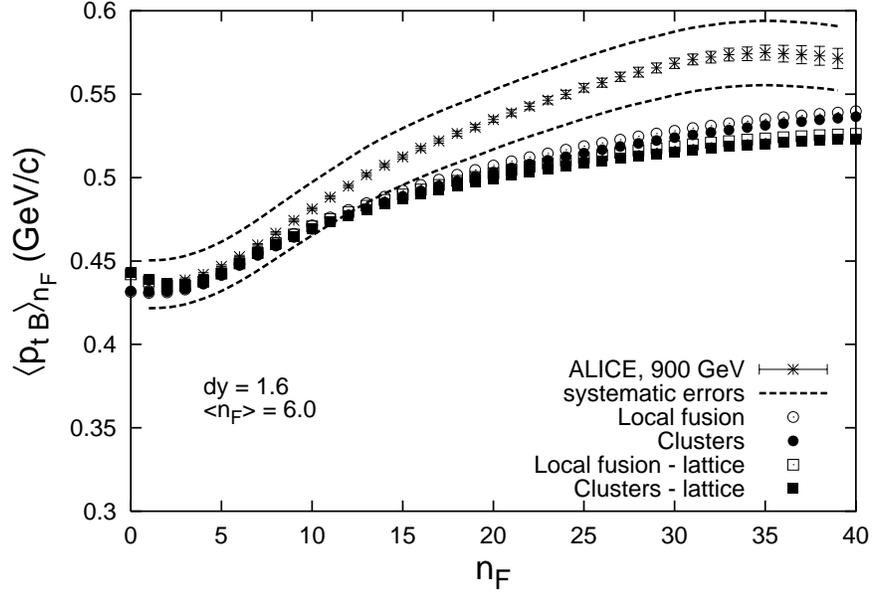,width=8cm,angle=-90}
\caption[dummy]{
The long-range correlation (LRC) function in $pp$ collisions between the multiplicity $\nF$ in the forward rapidity window and the
corresponding mean transverse momentum $\pB$ of $\nB$ charged particles in the backward window calculated
on the base of the formula (\ref{ptncor}) at the energy 900 GeV
for the forward and backward rapidity windows
of the same width $\DF=\DB=dy=1.6$, what corresponds to the mean multiplicity $\av\nF=\av\nB=\av n$=6.0 in the
window.  Notations of the points are the same as in Fig.\ref{fignn}.
Points $\times$ - the experimental data of the ALICE collaboration \cite{ALpt} on the correlation
between the multiplicity $n$ and the mean transverse momentum $p_t$ of charged particles in the same
pseudorapidity interval $\eta\in(-0.8, 0.8)$, $\Deta=1.6$ obtained in $pp$ collisions at 900 GeV.
}
\label{figptn}
\end{figure}
In Figs. \ref{fignn} and \ref{figptn} we present as an example the results of our calculations
of the LRC functions in $pp$ interactions
using the MC algorithm  based on the model described above.

In Fig.\ref{fignn} the results of calculations of LRC between multiplicities $\nF$ and $\nB$ in separated rapidity
windows on the base of the formula (\ref{nncor}) at the initial energies 63 and 900 GeV  are presented.
In both cases the rapidity width of the forward and backward windows is taken the same $\DF=\DB=dy=0.8$,
what corresponds to the mean multiplicities $\av\nF=\av\nB=\av n$=1.5 and 3.0 in such window at these energies.
We see in Fig.\ref{fignn} the considerable increase of the strength of the LRC between
multiplicities in separated rapidity windows.

In Fig.\ref{figptn} the results on the LRC between the multiplicity $\nF$ in the forward rapidity window and the
corresponding mean transverse momentum $\pB$ of $\nB$ charged particles in the backward window
on the base of the formula (\ref{ptncor}) at the initial energy 900 GeV  are presented.
The rapidity width of the forward and backward windows is taken the same $\DF=\DB=\Dy=1.6$,
what corresponds to the mean multiplicities $\av\nF=\av\nB=\av n$=6.0 in such window at this energy.
In Fig.\ref{figptn}  we also present the experimental data of the ALICE collaboration \cite{ALpt} on the correlation
between the multiplicity $n$ and the mean transverse momentum $p_t$ of charged particles in the same
pseudorapidity window $\eta\in(-0.8, 0.8)$, $\Deta=1.6$ obtained in $pp$ collisions at the energy 900 GeV.
The difference between calculated LRC function and the ALICE experimental data
at large multiplicities can be explained by
the contributions of additional short-range mechanisms in the case of the correlation
between the multiplicity $n$ and the mean transverse momentum $p_t$ in the same window,
which do not contribute in the case of LRC.

We see also in Figs. \ref{fignn} and \ref{figptn}  that the results of calculations of the LRC strength
in the frameworks of all four versions of SFM
(local fusion or cluster formation and their lattice analogs, see Figs.\ref{SFMver})
turn out to be very close to each other.
On the one hand, because of the principal technical differences in the
realization of these SFM versions, it indicates in favour of the reliability of the obtained results.
On the other hand, it obviously means that
one can not distinguish between versions of the SFM mechanisms
on the base of this comparison of the obtained results with the experimental data
and the calculations of other string fusion effects on the physical observables are needed.

\section{Conclusions}
The simple model which enables to take into account the effect of colour string fusion in $pp$ interactions
is suggested.  At the formulation of the model
we assume that the dependence of the average number of cut pomerons
on the impact parameter in a non-diffractive (ND) \textit{pp} collision
is gaussian (\ref{a12}) with the additional condition that we have at least one cut pomeron (\ref{a12a}).
We assume also that the event-by-event distribution of the number of cut pomerons
around this average value at fixed impact parameter $b$
is poissonian (\ref{a10}) with the same condition (\ref{a10a}).

It is shown that these two simple assumptions
after integration over impact parameter
 lead to the well known
formula for the cross-section $\sigma_N$ with $N$ cut-pomeron exchange
in a ND  \textit{pp} collision (\ref{sigN}), (\ref{R}),
which was obtained in the quasi-eikonal
and Regge approaches \cite{Ter-Mart73,AGK,Kaid84}.
This have enabled us to connect  the parameters of our model with  the parameters of the pomeron trajectory
and its couplings to hadrons (\ref{paramfix}).

The effects of the string fusion \cite{BP1,BP2} on the multiparticle production
were taken into account in the same way as it was done in the case
of $AA$ collisions \cite{EPJC04,YF1}.
At that the different version of the string fusion mechanism
(local fusion or cluster formation and their lattice analogs, see Fig.\ref{SFMver})
were considered.

On the base of the model the Monte-Carlo algorithm was developed and
the long-range correlation functions between multiplicities and between the average transverse momentum
and the multiplicity in \textit{pp} collisions at different energies were calculated.
It was found that the results of calculations of the long-range correlation (LRC) strength
in the frameworks of all four versions of the string fusion model (SFM)
turn out to be very close to each other
(see Figs. \ref{nncor} and \ref{ptncor}).


\begin{thebibliography}{99}
%
\bibitem{Capella0}
      A.~Capella and A.~Krzywicki, \emph{Phys. Rev. D} {\bf 18} (1978) 4120.
\bibitem{Capella1}
      A.~Capella, U.P.~Sukhatme, C.--I.~Tan and J.~Tran Thanh Van,
                  \emph{Phys. Lett. B} {\bf 81} (1979) 68; \emph{Phys. Rep.} {\bf 236} (1994) 225.
\bibitem{Kaidalov0}
      A.B.~Kaidalov, \emph{Phys. Lett. B} {\bf 116} (1982) 459.
\bibitem{Kaidalov1}
      A.B.~Kaidalov K.A.~Ter-Martirosyan, \emph{Phys. Lett. B} {\bf 117} (1982) 247.

\bibitem{BP1}
M.A.~Braun, C.~Pajares, \emph{Phys. Lett. B} {\bf 287} (1992) 154; \emph{Nucl. Phys. B} {\bf 390} (1993) 542.
\bibitem{BP2}
      N.S.~Amelin, M.A.~Braun, C.~Pajares, \emph{Phys. Lett. B} {\bf 306} (1993) 312;
      \emph{Z. Phys. C} {\bf 63} (1994) 507.
\bibitem{PRLperc}
      N.~Armesto, M.A.~Braun, E.G.~Ferreiro, C.~Pajares, \emph{Phys. Rev. Lett.} {\bf 77} (1996) 3736.

%
\bibitem{comparRHIC1}
M.A.~Braun, F.~del~Moral, C.~Pajares, \emph{Phys. Rev. C} {\bf 65} (2002) 024907.
\bibitem{comparRHIC2}
N.~Armesto, C.~Pajares, D.~Sousa, \emph{Phys. Lett. B} {\bf 527} (2002) 92.

\bibitem{PRL94}
      N.S.~Amelin, N.~Armesto, M.A.~Braun, E.G.~Ferreiro, C.~Pajares,
      \emph{Phys. Rev. Lett.} {\bf 73} (1994) 2813.
\bibitem{BP00}
      M.A.~Braun and C.~Pajares, \emph{Phys. Rev. Lett.} {\bf 85} (2000) 4864.
\bibitem{BPep00}
      M.A.~Braun and C.~Pajares, \emph{Eur. Phys. J. C} {\bf 16} (2000) 349.
\bibitem{BPV00}
     M.A.~Braun, C.~Pajares, V.V.~Vechernin, \emph{Phys. Lett. B} {\bf 493} (2000) 54 [{\tt hep-ph/0007241}].

\bibitem{AL_INT}
     M.A.~Braun, C.~Pajares, V.V.~Vechernin,
\emph{Forward-backward multiplicity correlations, low pt distributions in the central region
and the fusion of colour strings}
Internal Note/FMD, ALICE-INT-2001-16,
CERN, Geneva 2001.

\bibitem{Bol}
P.A.~Bolokhov, M.A.~Braun, G.A.~Feofilov,
V.P.~Kondratiev, V.V.~Vechernin,
\emph{Long-Range Forward-Backward Pt and Multiplicity
Correlations Studies in ALICE}
Internal Note/PHY, ALICE-INT-2002-20,
CERN, Geneva 2002.
\bibitem{PPR2}
ALICE collaboration, \emph{ALICE: Physics Performance Report, Volume II}, \emph{J. Phys. G} \textbf{32} (2006) 1295-2040
(Section: 6.5.15 - \emph{Long-range correlations}, p.1749-1751).
\bibitem{ALP}
N.~Armesto, L.~McLerran, C.~Pajares, \emph{Nucl. Phys. A} (2007)  781.

\bibitem{pp1}
C.~Albajar et al. (UA1 Collaboration), \emph{Nucl. Phys. B} {\bf 335} (1990) 261.
\bibitem{pp2}
F.~Abe et al. (CDF Collaboration), \emph{Phys. Rev. D} {\bf 61} (2000) 032001.
\bibitem{pp3}
T.~Alexopoulos et al. (E735 Collaboration), \emph{Phys. Rev. Lett.} 60 (1988) 1622;
\emph{Phys. Rev. D} {\bf 48} (1993) 984;
\emph{Phys. Lett. B} {\bf 336} (1994) 599.
\bibitem{pp5}
A.~Breakstone et al. (ABCDHW Collaboration), \emph{Phys. Lett. B} {\bf 132} (1983) 463.

\bibitem{Bialas76}
A.~Bialas, M.~Bleszynski, W.~Czyz, \emph{Nucl. Phys. B} {\bf 111} (1976) 461.
\bibitem{PRC11}
V.V.~Vechernin, H.S.~Nguyen, \emph{Phys. Rev. C} {\bf 84} (2011) 054909 [{\tt 1102.2582 [hep-ph]}].

\bibitem{Ter-Mart73}
K.A.~Ter-Martirosyan, \emph{Phys. Lett. B} {\bf 44} (1973) 377.
\bibitem{AGK}
V.A.~Abramovsky, V.N.~Gribov, O.V.~Kancheli, \emph{Yad. Fiz.} {\bf 18} (1973) 595.
\bibitem{Kaid84}
A.B.~Kaidalov, K.A.~Ter-Martirosyan, \emph{Yad. Fiz.} {\bf 39} (1984) 1545; \emph{ibid.} {\bf 40} (1984) 211.
\bibitem{KaidShab02}
G.H.~Arakelyan, A.~Capella, A.B.~Kaidalov, Yu.M.~Shabelski, \emph{Eur. Phys. J. C} {\bf 26} (2002) 81.
\bibitem{loc}
M.A.~Braun, C.~Pajares, \emph{Eur. Phys. J. C} {\bf 16} (2000) 349.
\bibitem{glob}
M.A.~Braun, F.~ del Moral, C.~Pajares, \emph{Phys. Rev. C} {\bf 65} (2002) 024907.
\bibitem{V1}
V.V.~Vechernin, R.S.~Kolevatov, \emph{Vestnik SPbU}, ser.4, no.2, (2004) 12 [{\tt hep-ph/0304295}].
\bibitem{V2}
V.V.~Vechernin, R.S.~Kolevatov, \emph{Vestnik SPbU}, ser.4, no.4, (2004) 11 [{\tt hep-ph/0305136}].
\bibitem{EPJC04}
M.A.~Braun, R.S.~Kolevatov, C.~Pajares, V.V.~Vechernin,
\emph{Eur. Phys. J. C} {\bf 32} (2004) 535 [{\tt hep-ph/0307056}].
\bibitem{YF1}
V.V.~Vechernin, R.S.~Kolevatov, \emph{Physics of Atomic Nuclei} {\bf 70} (2007) 1797.
\bibitem{YF2}
V.V.~Vechernin, R.S.~Kolevatov, \emph{Physics of Atomic Nuclei} {\bf 70} (2007) 1858.

\bibitem{Dub10}
V.V.~Vechernin, \emph{Long-Range Rapidity Correlations in the Model with Independent Emitters},
in proceedings of \emph{The XX International Baldin Seminar on High Energy Physics Problems},
vol.2, JINR, Dubna (2011) 10 [{\tt 1012.0214 [hep-ph]}].
\bibitem{CERN12}
V.V.~Vechernin, \emph{Correlations between multiplicities in rapidity and azimuthally separated windows},
{\tt 1210.7588 [hep-ph]}.

\bibitem{ALpt}
K. Aamodt et al. (ALICE collaboration),  \emph{Phys. Lett. B} \textbf{693} (2010) 53  [{\tt 1007.0719 [hep-ex]}].

\end{thebibliography}
\end{document}